\documentclass[twoside]{article}

\usepackage[preprint]{aistats2026}
\usepackage[authoryear,round]{natbib}

\usepackage{amsmath,amssymb,amsfonts}
\usepackage[colorlinks=true, citecolor=blue, linkcolor=blue, urlcolor=blue]{hyperref}
\usepackage{mathtools}

\newtheorem{theorem}{Theorem}[section]
\usepackage{xcolor} 
\newtheorem{lemma}{Lemma}
\newtheorem{proposition}{Proposition}

\newtheorem{assumption}{Assumption}
\usepackage{algorithm}
\usepackage{algpseudocode}
\usepackage{booktabs}
\usepackage{float}

\begin{document}

%

%

\twocolumn[


\aistatstitle{Learning Across Experiments and Time: Tackling Heterogeneity in A/B Testing}

\aistatsauthor{Xinran Li}

\aistatsaddress{University of Science and Technology of China \\ lixinran2022@mail.ustc.edu.cn } ]

\begin{abstract}

A/B testing plays a central role in data-driven product development, guiding launch decisions for new features and designs. However, treatment effect estimates are often noisy due to short horizons, early stopping, and slowly accumulating long-tail metrics, making early conclusions unreliable. A natural remedy is to pool information across related experiments, but naive pooling potentially fails: within experiments, treatment effects may evolve over time, so mixing early and late outcomes without accounting for nonstationarity induces bias; across experiments, heterogeneity in product, user population, or season dilutes the signal with unrelated noise. These issues highlight the need for pooling strategies that adapt to both temporal evolution and cross-experiment variability. To address these challenges, we propose a local empirical Bayes framework that adapts to both temporal and cross-experiment heterogeneity. Throughout an experiment's timeline, our method builds a tailored comparison set: time-aware within the experiment to respect nonstationarity, and context-aware across experiments to draw only from comparable counterparts. The estimator then borrows strength selectively from this set, producing stabilized treatment effect estimates that remain sensitive to both time dynamics and experimental context. Through theoretical analysis and empirical evaluation, we show that the proposed local pooling strategy consistently outperforms global pooling by reducing variance while avoiding bias. Our proposed framework enhances the reliability of A/B testing under practical constraints, thereby enabling more timely and informed decision-making.

\end{abstract}

\section{Introduction}
A/B~testing, commonly referred to as online controlled experimentation, has become a core tool for data-driven organizations to evaluate product and service changes, test novel ideas, and support evidence-based decision making. In its basic form, an A/B test randomly assigns users to a control (A) or treatment (B) group, enabling causal effects to be estimated by comparing outcomes between groups. Large-scale experimentation infrastructures are now deeply integrated into major technology companies, supporting tens of thousands of concurrent experiments and engaging millions of users each year \citep{Tang2010Overlapping,Gupta2019Challenges}. This widespread adoption underscores the role of A/B testing as both a driver of industrial innovation and a methodological cornerstone of empirical research in statistics and applied machine learning \citep{Kohavi2007Guide,Kohavi2020Book}.

While A/B testing has become the standard for large-scale decision making, a central challenge is to ensure that the estimated treatment effects (the differences in outcomes between treatment and control groups)are reliable. Organizations rely on these estimates to guide product launches, policy changes, and other high-stakes decisions. When estimates are unreliable, the credibility of experimentation is undermined and its value as a foundation for data-driven decision making is diminished \citep{Kohavi2020Book}. In practice, this challenge is amplified by platform constraints: readouts often rely on \emph{short horizons}, interim analyses involve \emph{early stopping}, and many \emph{long-tail metrics} accumulate slowly, making early conclusions noisy and potentially misleading. To evaluate and improve reliability, prior work has focused on mean squared error (MSE) of treatment effect estimators as the key evaluation criterion. MSE combines variance and squared bias into a single measure, directly capturing both the precision and accuracy of effect estimation \citep{Casella2001Statistical,Hastie2009Elements}. Reducing MSE is therefore essential for producing dependable experimental findings and for enhancing the reliability of online experimentation \citep{Johari2017Peeking,Kohavi2020Book}.

A wide range of methods has been developed specifically to reduce mean squared error (MSE) in A/B testing, including control variates and pre-post techniques \citep{deng2013cuped, xie2016improving, deng2017continuous}, regression adjustment \citep{lin2013regression}, and blocking or stratification designs \citep{imai2008misunderstandings}. Beyond these directions, Empirical Bayes (EB) shrinkage, as a cross-experiment approach, is distinctive in that it tackles high-variance regimes that within-experiment methods cannot adequately address, making it a particularly valuable strategy for reducing MSE \citep{james1961estimation, CarlinLouis2000Bayes, efron2010largescale, azevedo2019empirical, guo2020empirical}. EB has its roots in the James–Stein estimator and empirical Bayes theory \citep{efron2010largescale, CarlinLouis2000Bayes}, and it reduces variance by pooling information across related experiments, which makes estimates more reliable even when they would otherwise be highly noisy. Although shrinkage introduces some bias, the variance reduction often dominates, yielding a lower overall MSE. Owing to its effectiveness in reducing MSE, EB has become a standard baseline in both statistics and large-scale experimentation.

However, the effectiveness of classic empirical Bayes (EB) shrinkage hinges on a strong \emph{exchangeability} assumption, namely that all experiments are drawn from the same underlying effect distribution. In practice, this corresponds to shrinking every estimate toward a single global mean. Yet in large-scale experiments this assumption is rarely satisfied, as experiments often exhibit substantial heterogeneity stemming from two sources. \textbf{Within experiments}, treatment effects may evolve over time as user behavior and market conditions vary across seasons or holidays \citep{Wu2022Nonstationary,Deng2016Continuous}, and mixing early and late outcomes without accounting for \emph{nonstationarity} induces bias. \textbf{Across experiments}, heterogeneity arises from differences in product surfaces, user populations, and seasonal effects  \citep{Tang2010Overlapping,Kohavi2020Book}, and pooling across such types dilutes true signal with unrelated noise. In both cases, the systematic bias introduced by inappropriate pooling can offset, or even outweigh the variance reduction benefits. This limits MSE gains and may cause global EB to underperform simpler estimators. \textbf{These limitations} highlight a core constraint of global pooling and underscore the need for shrinkage strategies that respect heterogeneity in real-world experimentation.

To date, a fundamental question remains largely unanswered:
\textit{“Can EB methods be extended beyond global exchangeability to remain valid under real-world heterogeneity?”} This question motivates our work.

To address this challenge, we propose a Local Empirical Bayes (EB) shrinkage framework. The central idea is that similar experiments provide similar information. Rather than shrinking all estimates toward a single global mean, we adaptively identify for each target experiment a neighborhood of comparable experiments and perform shrinkage within this local group. Our design is \emph{time-aware within experiments}, respecting nonstationarity, and \emph{context-aware across experiments}, ensuring that only comparable counterparts contribute to shrinkage. This localized design extends the applicability of EB methods to heterogeneous settings. From a theoretical perspective, we establish that when treatment effects are heterogeneous, local EB achieves strictly lower overall mean squared error (MSE) than classical EB, as it retains the variance reduction of shrinkage while avoiding the systematic bias from inappropriate global pooling. We further validate our framework on a large-scale dataset of online experiments from ASOS, demonstrating its practical effectiveness in real-world applications.

Our contributions are threefold:
\paragraph{Methodology:}We propose a local empirical Bayes (EB) shrinkage framework that replaces global pooling with small, experiment-specific neighborhoods. We instantiate this idea with the \emph{Cross-fitting Stratified Hybrid Neighbors (CF-SHN)} method, which combines process features and outcome similarity to enable localized and selective information borrowing under nonstationarity and heterogeneity.
\paragraph{Theory:}We provide formal analysis showing that under treatment-effect heterogeneity, local EB achieves strictly lower mean squared error (MSE) than classical EB. The improvement arises because variance reduction is preserved while systematic bias from inappropriate global pooling is avoided. 
\paragraph{Empirics:}We validate our framework on a large-scale dataset from ASOS, covering 78 experiments and 24{,}153 snapshots. Our method reduces overall MSE by $27.2\%$ relative to the no-shrinkage baseline (compared to only $1.2\%$ for classical EB), and improves performance in $82.4\%$ of individual experiments, demonstrating both substantial and robust gains. 

The rest of the paper proceeds as follows. We review related work in Section~\ref{sec:related}, and introduce the classical EB baseline together with our evaluation metric in Section~\ref{sec:background}. Section~\ref{sec:method} details the proposed local EB framework, including the neighborhood selection method. Section~\ref{sec:theory} establishes theoretical guarantees, and Section~\ref{sec:simulation} demonstrates the empirical performance on large-scale experiments.

\section{Related Work}\label{sec:related}
Our work bridges two strands of literature: empirical Bayes methods for experimentation and localization techniques for handling heterogeneity. Although both have been extensively studied on their own, their integration in the context of online A/B testing, where heterogeneity arises both across and within experiments, has not been systematically explored.
\paragraph{Empirical Bayes in A/B Testing}
Empirical Bayes (EB) shrinkage has become a cornerstone of modern large-scale experimentation, valued for its ability to stabilize noisy treatment effect estimates by pooling information across tests. Rooted in the James–Stein estimator and formalized in EB theory \citep{james1961estimation, CarlinLouis2000Bayes, efron2010largescale}, the classical EB approach shrinks estimates toward a single global prior mean. This framework is now widely used as a baseline for mitigating the winner’s curse and stabilizing early readouts \citep{ azevedo2019empirical}. Over time, the literature has adapted EB to the realities of online experimentation, producing extensions such as spectral prior estimation \citep{guo2020empirical}, multivariate procedures \citep{Banerjee2020NEST}, covariate-powered EB \citep{IgnatiadisWager2019}, and methods addressing post-selection inference \citep{Deng2021PostSelection} and continuous monitoring \citep{Xu2022AMSET}.

However, despite these advances, a fundamental limitation remains: the assumption of a common effect distribution across all experiments. In practice, this exchangeability assumption is often violated due to temporal nonstationarity and cross-experiment heterogeneity \citep{Tang2010Overlapping, Kohavi2020Book, Wu2022Nonstationary, Deng2016Continuous}. As emphasized in applied studies, shrinking toward an irrelevant global mean can introduce systematic bias that offsets the variance reduction and erodes MSE gains. This concern has motivated recent efforts to detect heterogeneity \citep{Adam2024Heterogeneity} and, more broadly, to develop \textbf{localized} EB methods.

\paragraph{The Localization Frontier and Our Position}
The principle of localization, which adapts shrinkage to data substructures, has been explored in foundational statistical work \citep{efron2008microarrays}. Neighborhood-based EB \citep{ zhang1997empirical, jiang2010empirical} and covariate-assisted procedures in multiple testing and genomics \citep{stephens2017ash, ignatiadis2016data} illustrate the benefits of moving beyond a global mean. Yet, these approaches typically rely on static covariates and have seen little application in online experimentation. Importantly, the idea of local EB remains largely unexplored in A/B testing, where one must address dual-horizon heterogeneity by defining similarity \textbf{across experiments} in a context-aware way and \textbf{within experiments} in a time-aware way.

Our framework addresses this gap by integrating tools for quantifying experiment similarity. At the \textbf{process level}, we employ dynamic time warping (DTW) \citep{sakoe1978dtw, berndt1994using, keogh2005exact} to compare nonstationary arrival patterns. At the \textbf{outcome level}, we leverage meta-analytic techniques \citep{DerSimonian1986Meta, Hedges1985Statistical, Veroniki2016Methods} to aggregate effect estimates in a principled manner. By combining these perspectives, we construct experiment-specific neighborhoods that support a novel, fully localized EB shrinkage strategy tailored to the unique challenges of online experimentation.

\section{Preliminaries}\label{sec:background}

In this section, we review the classical empirical Bayes (EB) shrinkage framework and the mean squared error (MSE) criterion commonly used to evaluate treatment effect estimators in A/B testing. 

\subsection{Classical Empirical Bayes (EB) Shrinkage}
Consider a collection of $K$ experiments, indexed by $k=1,\dots,K$. 
Each experiment yields an unbiased effect estimate $y_k$ of the true effect $\theta_k$, with known sampling variance $v_k > 0$. 
The classical EB model assumes that all effects are exchangeable draws from a common Gaussian prior,
\[
\theta_k \sim \mathcal{N}(\mu,\tau^2), 
\qquad 
y_k \mid \theta_k \sim \mathcal{N}(\theta_k, v_k).
\]
The hyperparameters $(\mu,\tau^2)$ are estimated from the pooled data $\{(y_j,v_j)\}_{j=1}^K$ 
using standard EB procedures \citep{efron2010largescale, CarlinLouis2000Bayes}. 
This yields the shrinkage estimator
\[
\tilde{\theta}_k = (1-B_k)\mu + B_k y_k, 
\qquad 
B_k = \frac{\tau^2}{\tau^2 + v_k},
\]
where $B_k$ balances the noisy observation $y_k$ against the global mean $\mu$. 
This shrinkage reduces variance when $v_k$ is large and serves as a standard baseline in large-scale experimentation. 

However, its effectiveness rests on the exchangeability assumption that all experiments share a common distribution, an assumption often violated in practice due to heterogeneity. As a result, classical EB is not well-suited for heterogeneous online experimentation settings.

\subsection{Mean Squared Error (MSE) in A/B Testing}
In a standard A/B test, $n_T$ units are assigned to treatment and $n_C$ to control. Let $\bar{Y}_T$ and $\bar{Y}_C$ denote the sample means in the two groups, with population variances $\sigma_T^2$ and $\sigma_C^2$. 
The raw difference-in-means $y = \bar{Y}_T - \bar{Y}_C $ estimates the true effect
\[
\theta = \mathbb{E}[Y_T] - \mathbb{E}[Y_C],
\]
and under randomization satisfies $\mathbb{E}[y]=\theta$ with sampling variance
\[
v = \frac{\sigma_T^2}{n_T} + \frac{\sigma_C^2}{n_C}.
\]
Thus its mean squared error equals the variance, $\mathrm{MSE}(y)=v$, which provides a natural baseline.

For a general estimator $\hat{\theta}$ of $\theta$, however,
\[
\mathrm{MSE}(\hat{\theta}) 
= \mathbb{E}\!\left[(\hat{\theta}-\theta)^2\right]
= \mathrm{Var}(\hat{\theta}) + \mathrm{Bias}(\hat{\theta})^2,
\]
which we use as the primary performance criterion. This decomposition highlights the central trade-off in shrinkage: variance reduction is beneficial only when the bias induced by an imperfect target does not dominate. These considerations motivate the localized extension of classical EB introduced in Section~\ref{sec:method}.

\section{Local Empirical Bayes Framework}\label{sec:method}

This section proposes our local EB framework, which adapts shrinkage to heterogeneity by constructing tailored neighborhoods that are \emph{time-aware within experiments} to respect nonstationarity and \emph{context-aware across experiments} to ensure comparability.

\subsection{From Global to Local Shrinkage}

To address the limitations of global EB under heterogeneity, we propose a local EB framework whose key innovation is to replace the single global mean with an adaptively estimated, neighborhood-specific target.

For each experiment $k$, we form a neighborhood $N_k$ of comparable experiments and estimate local hyperparameters $(\hat{\mu}_k, \hat{\tau}_k^2)$ from this group. The estimator retains the canonical EB form but is tailored to the local context:
\[
\tilde{\theta}_k = (1-B_k)\hat{\mu}_k + B_k y_k, 
\qquad 
B_k = \frac{\hat{\tau}_k^2}{\hat{\tau}_k^2 + v_k}.
\]
Here $B_k$ balances variability within the neighborhood ($\hat{\tau}_k^2$) against the sampling noise of the current experiment ($v_k$), thereby shifting weight toward the local mean $\hat{\mu}_k$ when the neighborhood is homogeneous or the observation is noisy.

By localizing shrinkage targets, our framework preserves the variance reduction benefits of EB while mitigating the bias caused by inappropriate global pooling, making it better suited for heterogeneous experimentation environments.

\subsection{Local Similarity Metric for Experiments}

Next, to operationalize neighborhoods that are \emph{time-aware within experiments} and \emph{context-aware across experiments}, we define similarity along two complementary dimensions that directly target the sources of heterogeneity observed in practice. Each experiment $k$ is characterized by a feature vector
\[
\Phi_k = \big(\tilde{\lambda}_k(t), \log n_k\big), \quad 
n_k = \int_0^T \lambda_k(t) \, dt,
\]
where $\tilde{\lambda}_k(t)$ is its normalized temporal arrival pattern (capturing within-experiment nonstationarity) and $\log n_k$ is its overall traffic scale (capturing cross-experiment differences in magnitude). This design ensures that comparing arrival patterns across experiments makes neighborhoods, and hence the shrinkage, respect nonstationarity and avoid conflating early and late phases.

Similarity between experiments $i$ and $j$ is then quantified by a composite distance:
\[
D(\Phi_i,\Phi_j) 
= \rho \, \frac{d_{\mathrm{DTW}}(\tilde{\lambda}_i,\tilde{\lambda}_j)}{\operatorname{median}\{d_{\mathrm{DTW}}\}}
+ (1-\rho) \, \frac{|\log n_i - \log n_j|}{\operatorname{MAD}_{\log n}}.
\]
Here $d_{\mathrm{DTW}}$ denotes the \emph{dynamic time warping (DTW)} distance \citep{sakoe1978dtw}, which computes the minimal alignment cost between two temporal sequences, accommodating local shifts and stretches. The term $|\log n_i - \log n_j|$ measures differences in traffic volume, and the \emph{median absolute deviation (MAD)} is defined as $\operatorname{MAD}_{\log n} = \operatorname{median}_j \left( | \log n_j - \operatorname{median}_i(\log n_i) | \right)$
. Robust normalizations (median and MAD) ensure both components are comparable \citep{Huber2009Robust}, and the weight $\rho \in (0,1)$ balances shape versus scale.

By construction, this metric ensures that neighborhoods are simultaneously adapted to temporal nonstationarity and contextual heterogeneity, directly implementing our dual-horizon adaptation principle. \emph{Further implementation details, including temporal alignment, hyperparameter tuning, and computational complexity, are provided in Appendix~\ref{app1}.}

\subsection{Neighborhood Selection with CF-SHN}

Having established a principled similarity metric, we now address the core challenge of neighborhood selection: when $y_k$ is used for both neighbor selection and EB shrinkage, its noise is double-counted, inducing bias. To break this statistical dependence, we employ \emph{cross-fitting} \citep{Wager2018}, which produces proxy estimates $\hat{\mu}_j^{(-)}$ that are independent of the target experiment’s noise. These out-of-fold estimates provide denoised signals for neighborhood construction, ensuring the validity of subsequent inference.

We then introduce the \emph{Cross-Fitted Stratified Hybrid Neighbors (CF-SHN)} method, which forms the operational core of our local EB framework. CF-SHN constructs neighborhoods in two stages: (A) a \textbf{process-based} filtering step that ensures \emph{time-awareness and scale-awareness} by selecting a broad candidate set of experiments with similar dynamics and volumes, addressing temporal nonstationarity and scale differences; and (B) an \textbf{outcome-based} refinement step that ensures \emph{context-awareness} by selecting experiments with similar denoised outcomes, thereby addressing cross-experiment heterogeneity (Algorithm~\ref{alg:cfshn}, lines~\ref{alg:cfshn:delta}--\ref{alg:cfshn:selectq}). Final estimation uses \emph{restricted maximum likelihood (REML)} \citep{Raudenbush2002} to fit the local hierarchical model, which is a standard technique for unbiased variance-component estimation in linear mixed models (Algorithm~\ref{alg:cfshn}, line~\ref{alg:cfshn:reml}). The complete procedure is summarized in Algorithm~\ref{alg:cfshn}.

\begin{algorithm}[h]
\caption{Cross-Fitted Stratified Hybrid Neighbors (CF-SHN)}
\label{alg:cfshn}
\begin{algorithmic}[1]
\State \textbf{Input:} Target experiment $k$ with process features $\Phi_k$, out-of-fold predictions $\hat{\mu}^{(-)}$, outcomes $y_j$, variances $v_j$; parameters $M_0$ (candidate set size), $q$ (neighborhood size), distance metric $D(\cdot,\cdot)$
\State \textbf{Output:} Neighborhood $N_k$ and local EB estimate $\tilde{\theta}_k$

\State \textbf{Stage 1: Process filtering}
\State \label{alg:cfshn:stage1} Compute $D(\Phi_k,\Phi_j)$ for all $j \neq k$
\State Select candidate set $C_k$ of size $M_0$ with the smallest distances

\State \textbf{Stage 2: Outcome refinement (cross-fitted)}
\State \label{alg:cfshn:delta} Within $C_k$, compute $\Delta_j = |\hat{\mu}_k^{(-)}-\hat{\mu}_j^{(-)}|$
\State \label{alg:cfshn:selectq} Select the $q$ experiments with the smallest $\Delta_j$ to form $N_k$

\State \textbf{Stage 3: Local EB estimation}
\State \label{alg:cfshn:reml} Fit $y_j \sim \mathcal N(\mu, \tau^2+v_j)$ for $j\in N_k$ via REML
\State Obtain $(\hat{\mu}_k,\hat{\tau}_k^2)$ and compute $\tilde{\theta}_k \gets (1-B_k)\hat{\mu}_k + B_k y_k$, with $B_k = \hat{\tau}_k^2/(\hat{\tau}_k^2+v_k)$

\State \Return $N_k$, $\tilde{\theta}_k$
\end{algorithmic}
\end{algorithm}

As outlined in Algorithm~\ref{alg:cfshn}, this integrated procedure guarantees that neighborhoods are both \emph{process-homogeneous and outcome-comparable}, enabling the dual \emph{time-aware and context-aware} adaptation that classical EB lacks. The two-stage design, safeguarded by cross-fitting, directly instantiates our principle of dual-horizon adaptation while rigorously avoiding noise reuse. The parameters $\rho$, $M_0$, and $q$ govern the trade-off between localization and reliability, with sensitivity analyses reported in the simulation study and the Supplement. This construction also provides the methodological basis for our theoretical analysis in Section~\ref{sec:theory}.

\section{MSE Dominance of Local Shrinkage}\label{sec:theory}

This section establishes the theoretical advantages of local EB over classical EB under treatment-effect heterogeneity. We show that local EB achieves strictly lower overall mean squared error (MSE) by replacing the global shrinkage target with a data-adaptive local one. Our analysis begins with oracle properties and then extends to plug-in estimators, providing a rigorous foundation for the empirical gains in Section~\ref{sec:simulation}.

\subsection{Formalizing Heterogeneity}
First of all, to analyze the limitations of classical EB and motivate our approach, we extend the standard setup in Section~\ref{sec:background} to incorporate latent heterogeneity. Each experiment $k$ is assumed to belong to a latent type $z_k \in \mathcal{Z}$, drawn independently from a categorical distribution $\Pr(z_k=z)=\pi_z$ over $z \in \mathcal{Z}$. 
Conditional on its type, the true effect follows
\[
\theta_k \mid z_k=z \sim N(\mu_z,\tau_z^2),
\]
while the observation model remains as in Section~\ref{sec:background}, 
namely $y_k \mid \theta_k \sim N(\theta_k,v_k)$. 
Marginalizing over types then yields a Gaussian mixture prior \citep{McLachlan2000Finite}.

\[
\theta_k \sim \sum_{z\in\mathcal Z}\pi_z N(\mu_z,\tau_z^2),
\]
with mixture mean
\[
\mu_{\mathrm{mix}}=\sum_{z\in\mathcal Z}\pi_z\mu_z = \mathbb{E}[\mu_{z_k}].
\]
Classical EB shrinks toward this global center, effectively assuming all experiments arise from a single $N(\mu_{\mathrm{mix}},\tau^2)$ prior.
Thus, when the type mean $\mu_{z_k}$ deviates substantially from the mixture mean $\mu_{\mathrm{mix}}$, shrinkage toward $\mu_{\mathrm{mix}}$ can introduce large systematic bias.

To mitigate this issue, our local EB framework adapts the shrinkage target. For each experiment, we leverage observable features $X_k=(\Phi_k,\hat\mu^{(-)}_k)$, 
where $\Phi_k$ captures process descriptors and $\hat\mu^{(-)}_k$ is a cross-fitted pilot estimate, 
to construct the oracle local target
\[
\mu_{\mathrm{loc}}(X_k) = \mathbb{E}[\mu_{z_k}\mid X_k].
\]
We refer to this as an \emph{oracle}  \citep{Fan1993LocalLinear, Tsybakov2009Introduction} target because it relies on the true conditional distribution of $\mu_{z_k}$ given $X_k$, which is not available in practice but provides the ideal benchmark for our analysis. By definition, $\mu_{\mathrm{loc}}(X_k)$ is the closest feature-informed estimate of the type-specific mean. Intuitively, whenever features are informative about $z_k$, it lies closer to $\mu_{z_k}$ than the global mean does. We next formalize this intuition and quantify its consequences for MSE.

\emph{Our theoretical analysis focuses on feature-informed (context-aware) shrinkage, while the time-aware component enters through the construction of features $\Phi_k$ in practice. In addition, throughout this section, $\mathrm{MSE}$ denotes the Bayes risk (i.e., the expectation over both the latent prior and the sampling distribution).}

\subsection{Regularity Conditions}
Next, to establish our theoretical results, we begin by introducing a set of mild but essential assumptions. 

We first require that the latent types are meaningfully distinct, because otherwise local adaptation cannot improve upon classical EB.
\begin{assumption}[Heterogeneity]\label{assump1}
Formally, letting $\mu_{z_k}$ denote the mean associated with latent type $z_k \sim \pi$, we assume
\[
\mathrm{Var}(\mu_{z_k}) > 0.
\]
\end{assumption}
In the degenerate case $\mathrm{Var}(\mu_{z_k})=0$, all types share the same mean, and classical EB is already optimal. Building on this, we formalize when local features provide useful information about the latent type.

\begin{proposition}[Informative local target]\label{prop:informative}
Under Assumption~\ref{assump1}, if $X_k$ is not independent of $z_k$, then
\[
\mathrm{Var}\!\big(\mathbb{E}[\mu_{z_k}\mid X_k]\big) > 0.
\] 
\end{proposition}

Proposition~\ref{prop:informative} formalizes the key insight that whenever $X_k$ carries even partial information about the latent type, the oracle local target $\mu_{\mathrm{loc}}(X_k)$ varies across experiments and, in expectation, lies closer to the true mean $\mu_{z_k}$ than the uninformative global mean $\mu_{\mathrm{mix}}$. This is made precise by the inequality
\[
\mathbb{E}\!\left[(\mu_{\mathrm{loc}}(X_k)-\mu_{z_k})^2\right] 
< \mathbb{E}\!\left[(\mu_{\mathrm{mix}}-\mu_{z_k})^2\right],
\]
where the expectation is taken over $(z_k,X_k)$.

Next, we impose an assumption to rule out spurious improvements from reusing the same noise in both neighborhood construction and estimation. 

\begin{assumption}[Cross-fitting independence]\label{assump:cross-fitting}
For each target experiment $k$, the neighborhood $N_k$ and local hyperparameter estimates $(\hat\mu_k,\hat\tau_k^2)$ 
are constructed using data independent of the noise $\varepsilon_k = y_k - \theta_k$. Formally,
\[
(N_k, \hat\mu_k, \hat\tau_k^2) \;\perp\; \varepsilon_k \;\mid\; z_k.
\]
\end{assumption}

This assumption is satisfied by the cross-fitting protocol of \citet{Wager2018}, as implemented in Section~\ref{sec:method}. 
Cross-fitting separates selection from estimation, preventing adaptive overfitting and ensuring that performance gains reflect genuine information in the features rather than artifacts of noise reuse.

Finally, we impose standard boundedness conditions to exclude degenerate cases and keep shrinkage weights well behaved.  
\begin{assumption}[Bounded variances]\label{assump:bounded-variances}
There exist constants $0 < v_{\min} \leq v_k \leq v_{\max} < \infty$ for all sampling variances, and $0 \leq \tau_z^2 \leq \tau_{\max}^2 < \infty$ for all type variances $z \in \mathcal{Z}$.
\end{assumption}
This assumption ensures that shrinkage weights $B_k$ are uniformly bounded away from 0 and 1, excluding degenerate cases where the MSE improvement from a better target would be negligible. Such boundedness  \citep{Lehmann1998Theory, Casella1985Estimation} is standard in shrinkage analysis and ensures a fair comparison between local and classical EB.

These assumptions are mild and satisfied in practice, as heterogeneity is intrinsic to online experiments, cross-fitting is standard in our method, and bounded variances are routine. With these in place, we are ready to analyze the MSE of shrinkage estimators and isolate how the choice of target drives performance.

\subsection{Dominance Results}

With these regularity conditions in place, we now establish our main results: under heterogeneity with informative features, local EB achieves strictly lower overall MSE than its global counterpart. We begin by decomposing the risk of shrinkage estimators, which highlights the pivotal role of the shrinkage target.

\begin{lemma}[Risk decomposition of EB]\label{lem:decomp}
For any fixed shrinkage center $\tilde{\mu}$ and weight $B \in [0, 1]$, the Bayes risk conditional on $z_k$ is
\[
\mathrm{MSE}(B, \tilde{\mu}\mid z_k) 
= (1 - B)^2 \left\{\tau_{z_k}^2 + (\tilde{\mu} - \mu_{z_k})^2\right\} + B^2 v_k.
\]
\end{lemma}

This decomposition shows that the MSE comprises two components: the first term captures the intrinsic variability of $\theta_k$ within its type, and the second captures the sampling noise. While the weight $B$ balances these contributions, the decisive factor is the alignment between the shrinkage center $\tilde{\mu}$ and the true type mean $\mu_{z_k}$. 

This observation motivates a comparison of two oracle strategies. The \emph{global oracle} shrinks toward the mixture mean $\mu_{\mathrm{mix}}$, while the \emph{local oracle} shrinks toward $\mu_{\mathrm{loc}}(X_k)$. Both leverage the same variance reduction mechanism, but their expected squared bias, $\mathbb{E}[(\tilde{\mu} - \mu_{z_k})^2]$, differs fundamentally.

\begin{theorem}[Oracle dominance of local EB]\label{thm:oracle}
Under Assumptions~\ref{assump1}--\ref{assump:bounded-variances} and Proposition~\ref{prop:informative}, the local shrinkage center $\mu_{\mathrm{loc}}(X_k) = \mathbb{E}[\mu_{z_k} \mid X_k]$ is, in expectation, closer to the true type mean than the global center $\mu_{\mathrm{mix}} = \mathbb{E}[\mu_{z_k}]$. Consequently,
\[
\mathbb{E}[\mathrm{MSE}_{\mathrm{loc}}] \;<\; \mathbb{E}[\mathrm{MSE}_{\mathrm{glob}}].
\]
\end{theorem}

Theorem~\ref{thm:oracle} shows that under heterogeneity, feature-informed shrinkage strictly improves upon the global alternative. This result highlights the core benefit of localization. By adapting to latent structure through features, it achieves a superior bias–variance trade-off in the oracle setting.

Having established the oracle case, we now consider the practical setting where hyperparameters $(\mu, \tau^2)$ must be estimated from data. This raises the question of whether the oracle advantage persists with plug-in estimators. To ensure that it does, we assume that cross-fitted centers are unbiased and that local centers are no harder to estimate than global ones. Together with Proposition~\ref{prop:informative}, which guarantees a non-degenerate oracle gap, these conditions suffice to establish finite-sample dominance. 

\begin{theorem}[Plug-in dominance of local EB]\label{thm:plugin}
In addition to Assumptions~\ref{assump1}--\ref{assump:bounded-variances} and Proposition~\ref{prop:informative}, suppose the unbiasedness and no-harder-estimation conditions hold. Then
\[
\mathbb{E}\!\left[\mathrm{MSE}^{\mathrm{plug}}_{\mathrm{loc}}\right]
\;<\;
\mathbb{E}\!\left[\mathrm{MSE}^{\mathrm{plug}}_{\mathrm{glob}}\right],
\]
where for $j \in \{\mathrm{loc}, \mathrm{glob}\}$,
\[
\mathrm{MSE}^{\mathrm{plug}}_j
= (1-\hat{B}_j)^2 \left\{\tau_{z_k}^2 + (\hat{\mu}_j - \mu_{z_k})^2\right\} + \hat{B}_j^2 v_k.
\]
\end{theorem}

Theorem~\ref{thm:plugin} confirms that the advantage of local EB extends beyond the oracle regime. When estimation error is controlled, the feature-informed target remains systematically closer to the truth, ensuring that the reduction in MSE carries over to practical implementations.

\textbf{Discussion} Our results show that localization improves EB by achieving a better bias–variance trade-off, thereby reducing MSE whenever features are informative. The cross-fitting protocol ensures that this reduction reflects genuine information rather than artifacts of noise reuse. Although we rely on bounded-variance assumptions, which may be relaxed in future work, our analysis establishes that local EB systematically dominates classical EB in MSE, providing a solid foundation for the empirical evidence in Section~\ref{sec:simulation}. All proofs are provided in Appendix~\ref{app:proofs}.

\section{Simulation Study}\label{sec:simulation}

We complement our theoretical analysis with an empirical evaluation on a semi-synthetic testbed, which mimics real-world experimentation dynamics while providing access to approximate ground truth. Our goal is to assess whether local EB methods, and in particular our proposed CF-SHN, consistently deliver MSE improvements under heterogeneous conditions.

\subsection{Motivation and Dataset}
We base our study on the \textbf{ASOS Digital Experiments Dataset} \citep{asosdata}, a public collection of 78 large-scale A/B tests from a global fashion e-commerce platform. The dataset includes 24,153 \emph{snapshots} recorded at daily or 12-hour intervals during 2019–2020. Each snapshot reports aggregated treatment and control sample sizes, means, and variances, which are sufficient to reconstruct effect estimates and their sampling variances, although user-level data are unavailable. The dataset exhibits substantial heterogeneity, including weekday–weekend cycles and large differences in traffic scale.

Because only summary statistics are available, the true effects $\theta_k$ cannot be directly observed. To address this, we construct a semi-synthetic testbed that preserves observed heterogeneity while supplying approximate ground truth, thereby enabling a realistic yet controlled evaluation of shrinkage strategies. The details of the generative modeling are described in the next subsection. For concreteness, we focus on \emph{Metric~2} in this dataset, a continuous anonymized KPI that serves as a representative benchmark.

\subsection{Simulation Setup}
\paragraph{Generative model and evaluation}
We fit a generative model to the ASOS snapshots, modeling traffic as a nonhomogeneous Poisson process with shape--scale decomposition 
\[
\lambda_k(t) = n_k f_k(t),
\]
where $f_k(t)$ captures temporal patterns and $n_k$ the total volume. Given snapshot times $t_0 < \cdots < t_M$, we approximate the intensity as constant within each interval $[t_{i-1},t_i)$ with rate $\bar\lambda_i$ proportional to the observed count increment. The normalized shape function is
\[
f(t) = \frac{\bar\lambda_i}{\sum_j \bar\lambda_j h_j}, 
\quad t \in [t_{i-1},t_i), 
\quad h_i = t_i - t_{i-1},
\]
which ensures $\int f(t)\,dt = 1$ and $n_k = \sum_j \bar\lambda_j h_j$. Treatment and control groups are split evenly, and outcomes are modeled as Gaussian. Bootstrap replicates yield difference-in-means estimates $y_k$ and variances $v_k$.

Since the true effects $\theta_k$ are unobserved, we approximate reference effects $\theta_k^*$ by long-horizon aggregation. Method performance is evaluated by mean squared error relative to this reference,
\[
\mathrm{MSE}_m = \frac{1}{N}\sum_{k=1}^N \big(\hat{\theta}_{k,m} - \theta_k^*\big)^2,
\]
reported both in absolute terms and as reduction relative to the raw estimator. 

\paragraph{Evaluated Methods}
To compare different methods, we evaluate five estimators: 
(A) \textbf{No shrinkage}, raw difference-in-means; (B) \textbf{Classical EB}, pooling all experiments to a global prior; (C) \textbf{Outcome-only}, neighborhoods by pilot outcomes; (D) \textbf{Process-only}, neighborhoods by DTW similarity of arrival curves and traffic scale; and (E) \textbf{CF-SHN (ours)}, combining process filtering with outcome refinement.  

All local methods use $K=5$-fold cross-fitting and are evaluated across neighborhood sizes $q \in \{6,8,\dots,20\}$.
For CF-SHN, we set the candidate set size $M_0 = 30$ and the shape-scale weight $\rho = 0.75$ based on sensitivity analysis. Results aggregate $B=1000$ bootstrap replicates, reporting MSE reduction versus the raw estimator and win-rate (fraction of experiments where shrinkage improves upon raw). Sensitivity analyses for hyperparameters and NHPP modeling details are deferred to the Supplement.

\subsection{Empirical Findings}
\begin{figure}[t]
    \centering
    \includegraphics[width=\linewidth]{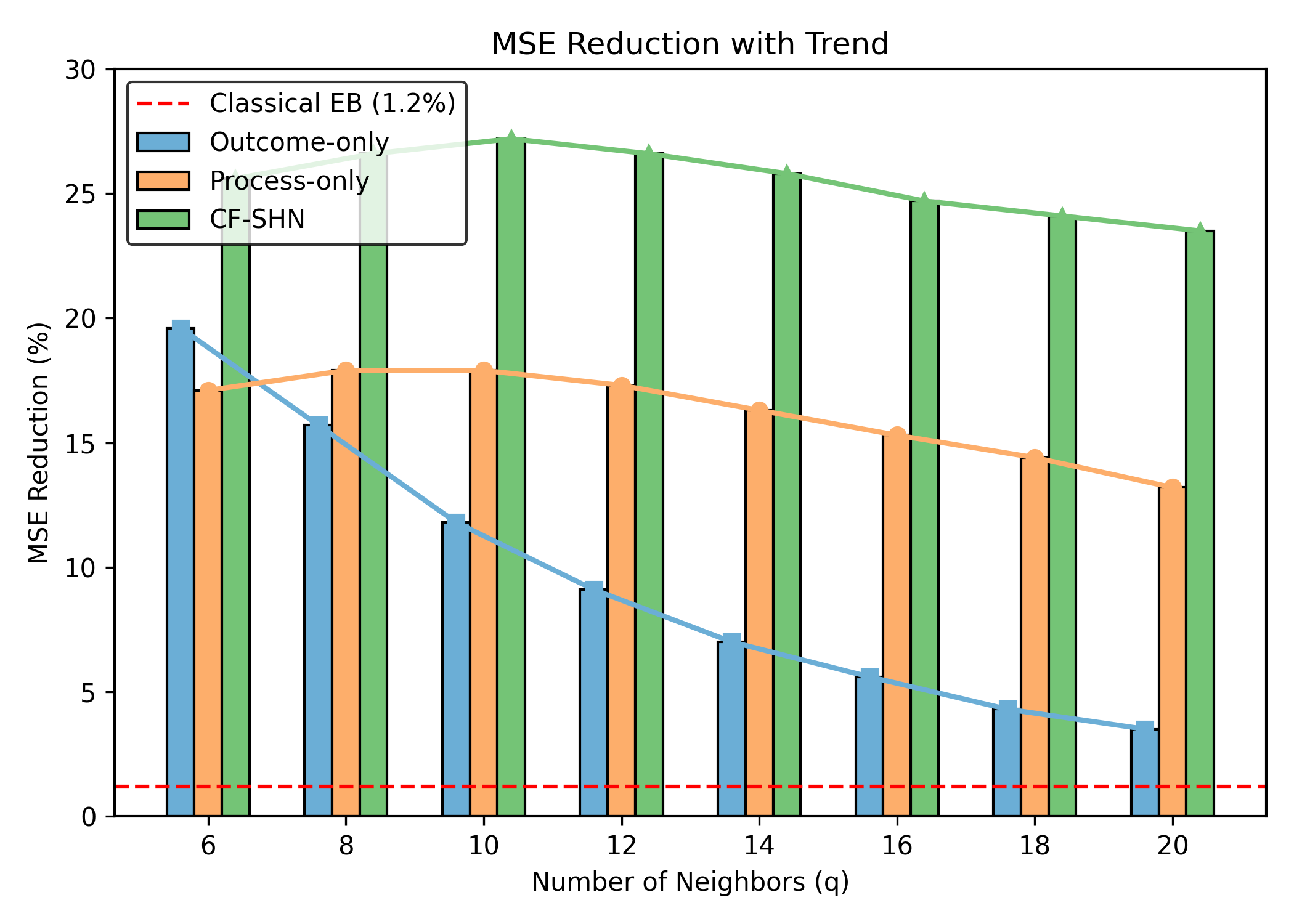}
    \caption{MSE reduction relative to the raw estimator on simulated ASOS experiments. 
Bars show average reductions and lines trace trends as $q$ increases. 
No shrinkage (0, not shown), classical EB, and our CF-SHN are included alongside local baselines.}
    \label{fig:mse_reduction}
\end{figure}

Figure~\ref{fig:mse_reduction} and Table~\ref{tab:mse_full} in Appendix summarize the performance of all methods.  
\textbf{Classical EB} delivers only negligible improvement, while local EB methods yield much larger gains. 
The \textbf{Outcome-only} approach starts strong but deteriorates quickly as $q$ increases, reflecting its sensitivity to neighborhood size. 
The \textbf{Process-only} method, in contrast, is more stable but achieves only moderate reductions. 
By comparison, \textbf{CF-SHN} consistently ranks first across all neighborhood sizes. 
It peaks at over $27\%$ reduction with win-rates above $80\%$, and even at larger $q$ values where other methods degrade, CF-SHN sustains over $23\%$ gains. 

Taken together, these results confirm that combining process similarity with outcome refinement under cross-fitting yields both higher accuracy and greater stability, establishing it as a reliable and principled solution for modern A/B testing under heterogeneity and reinforcing the theoretical advantages established in Section~\ref{sec:theory}.

\paragraph{Limitations and Future Work} 
Our simulation is limited to a single dataset and semi-synthetic design, so validation on diverse platforms remains important. 
The reliance on Poisson and Gaussian modeling may also miss more complex dynamics. 
Future work will extend CF-SHN to richer settings, including user-level data and real-time experimentation.

\section{Conclusion}
In this paper, we proposed a local empirical Bayes framework that adapts shrinkage to heterogeneity by constructing \emph{time-aware} and \emph{context-aware} neighborhoods. Through this approach, our theoretical analysis establishes the strict MSE dominance of local EB over its classical counterpart, while empirical results on large-scale ASOS experiments demonstrate substantial and robust improvements. Together, these advances position CF-SHN as a principled and reliable method for modern A/B testing, enhancing the credibility of treatment effect estimation in heterogeneous environments and enabling more confident, data-driven decision-making.

\bibliographystyle{plainnat}
\bibliography{references}

\section*{Checklist}

\begin{enumerate}

  \item For all models and algorithms presented, check if you include:
  \begin{enumerate}
    \item A clear description of the mathematical setting, assumptions, algorithm, and/or model. [Yes] (See Sections~\ref{sec:method} and \ref{sec:theory})
    \item An analysis of the properties and complexity (time, space, sample size) of any algorithm. [Yes] (Algorithm~\ref{alg:cfshn}, complexity in Appendix~\ref{app1})
    \item (Optional) Anonymized source code, with specification of all dependencies, including external libraries. [Not Applicable]
  \end{enumerate}

  \item For any theoretical claim, check if you include:
  \begin{enumerate}
    \item Statements of the full set of assumptions of all theoretical results. [Yes] (Assumptions~\ref{assump1}--\ref{assump:bounded-variances})
    \item Complete proofs of all theoretical results. [Yes] (Appendix~\ref{app:proofs})
    \item Clear explanations of any assumptions. [Yes] (Section~\ref{sec:theory})
  \end{enumerate}

  \item For all figures and tables that present empirical results, check if you include:
  \begin{enumerate}
    \item The code, data, and instructions needed to reproduce the main experimental results (either in the supplemental material or as a URL). [Yes] (Dataset \citep{asosdata} and simulation setup in Section~\ref{sec:simulation})
    \item All the training details (e.g., data splits, hyperparameters, how they were chosen). [Yes] (Section~\ref{sec:simulation} and Appendix~\ref{app1})
    \item A clear definition of the specific measure or statistics and error bars (e.g., with respect to the random seed after running experiments multiple times). [Yes] (MSE definition in Section~\ref{sec:background}, bootstrap details in Section~\ref{sec:simulation})
    \item A description of the computing infrastructure used. [Yes] (Simulation runtime and DTW computation described in Appendix~\ref{app1})
  \end{enumerate}

  \item If you are using existing assets (e.g., code, data, models) or curating/releasing new assets, check if you include:
  \begin{enumerate}
    \item Citations of the creator If your work uses existing assets. [Yes] (ASOS dataset \citep{asosdata})
    \item The license information of the assets, if applicable. [Not Applicable] (Dataset is public and open access)
    \item New assets either in the supplemental material or as a URL, if applicable. [Not Applicable]
    \item Information about consent from data providers/curators. [Not Applicable]
    \item Discussion of sensible content if applicable, e.g., personally identifiable information or offensive content. [Not Applicable]
  \end{enumerate}

  \item If you used crowdsourcing or conducted research with human subjects, check if you include:
  \begin{enumerate}
    \item The full text of instructions given to participants and screenshots. [Not Applicable]
    \item Descriptions of potential participant risks, with links to Institutional Review Board (IRB) approvals if applicable. [Not Applicable]
    \item The estimated hourly wage paid to participants and the total amount spent on participant compensation. [Not Applicable]
  \end{enumerate}

\end{enumerate}

\clearpage
\appendix
\onecolumn
\aistatstitle{Appendix}

\section{Process Features and Similarity Metric}\label{app1}

This section provides implementation details for constructing process-level features and computing DTW-based distances. 
These technical components are not essential for understanding the main framework but are crucial for ensuring robustness and reproducibility in practice.

\subsection{Temporal Normalization}

All experiments are observed over a fixed horizon $[0, T]$, where $T$ is chosen to fully cover the duration of each experiment. 
To compare experiments with different raw lengths, we rescale time to the unit interval $[0, 1]$. 
Let $s = t / T$ denote normalized time. 
The rescaled rate function is defined as
\[
\lambda_k^{\star}(s) = T \cdot \lambda_k(Ts),
\]
so that the total number of arrivals for experiment $k$ is
\[
n_k = \int_0^T \lambda_k(t)\,dt = \int_0^1 \lambda_k^{\star}(s)\,ds.
\]
We then normalize the rate function to obtain a shape-only curve:
\[
\tilde{\lambda}_k(s) = \frac{\lambda_k^{\star}(s)}{n_k} = \frac{T \cdot \lambda_k(Ts)}{n_k}, 
\quad \int_0^1 \tilde{\lambda}_k(s)\,ds = 1.
\]
This normalization removes scale effects and isolates the temporal dynamics of each experiment.

\subsection{Smoothing and Discretization}

Empirical arrival counts often contain high-frequency noise due to random variation in Poisson increments. 
To stabilize comparisons, we apply mild smoothing:
\begin{enumerate}
    \item For discrete arrival counts, we convolve the data with a Gaussian kernel whose bandwidth is proportional to $T^{1/4}$, balancing variance reduction and bias.
    \item Alternatively, we apply a moving-average filter with a fixed window (e.g., five minutes) to dampen sharp fluctuations.
\end{enumerate}
Sensitivity analysis confirms that the choice of smoothing method does not materially affect results over a broad range of bandwidths. 
For the experiments presented in this paper, we use Gaussian kernel smoothing with a fixed bandwidth $h=0.04$ after normalization to $[0,1]$.  
After discretization, each normalized curve $\tilde{\lambda}_k(s)$ is represented as a sequence of $L = 500$ equally spaced values.  

\subsection{Dynamic Time Warping Distance}

The DTW distance $d_{\mathrm{DTW}}(\tilde{\lambda}_i, \tilde{\lambda}_j)$ aligns two normalized sequences by allowing local stretching and compression along the time axis. 
Formally, it solves a dynamic programming problem of complexity $O(L^2)$, finding the minimum cumulative cost over all monotone warping paths. 
We adopt the standard squared-error cost between sequence elements:
\[
c(x_m, y_n) = (x_m - y_n)^2,
\]
where $x_m$ and $y_n$ denote the respective elements of the two sequences being compared.
To improve efficiency, we constrain the warping path using a Sakoe–Chiba band of width $\alpha L$ with $\alpha = 0.1$, 
which limits misalignment to $\pm 10\%$ of the sequence length. 
This constraint reduces runtime while preserving sufficient flexibility for local temporal shifts.

\subsection{Normalization and Computational Details}

The raw DTW distances and log-scale differences may have different magnitudes. 
To ensure comparability, we normalize both components using robust scale statistics:
\[
\mathrm{medDTW} = \mathrm{median}\{d_{\mathrm{DTW}}(\tilde{\lambda}_i, \tilde{\lambda}_j)\}, 
\quad
\mathrm{MAD}_{\log n} = \mathrm{median}_j \Big( |\log n_j - \mathrm{median}_i(\log n_i)| \Big).
\]
These statistics are computed once over all experiment pairs and remain fixed throughout the analysis. 
This normalization prevents any single term from dominating and makes the tuning parameter $\rho \in (0, 1)$ interpretable as a trade-off weight between the temporal shape and scale components.

The full pairwise DTW computation scales as $O(\alpha L^2)$.  
With $L=500$ and $\alpha=0.1$, computing all $78\times77/2$ pairwise DTW distances on the ASOS dataset 
takes less than 10 minutes on a MacBook~Pro (M2, 16~GB~RAM).  
The complete simulation and estimation pipeline, including $B=1000$ bootstrap replicates, 
local EB fitting, and CF-SHN neighborhood construction, requires about one hour of wall-clock time.  
Approximate variants such as \texttt{FastDTW}~0.4.0 yield nearly identical distances and 
can further reduce the DTW phase to under three minutes.

\paragraph{Reproducibility Note}
All computations were performed in Python~3.11 using standard packages (\texttt{numpy}, \texttt{scipy}, and \texttt{fastdtw}).  
Random seeds are fixed for all data resampling and DTW tie-breaking to ensure determinism.  
When the candidate set size $M_0<q$, we set $q\leftarrow M_0$.  
If REML estimation fails to converge, the method falls back to maximum likelihood (ML) and clips 
$\hat{\tau}^2 \ge 10^{-10}$.  
Cross-fitting folds are stratified by experiment size to maintain balanced allocation.  
These implementation details ensure stability and reproducibility across runs.

\section{Proof of Theory}\label{app:proofs}
This section provides detailed proofs of Proposition~1, Lemma~1, 
and Theorems~5.1–5.2 presented in Section~5 of the main paper.

\subsection{Proof of Proposition~1}

By the law of total variance,
\[
\mathrm{Var}(\mu_{z_k}) 
= \mathbb{E}\!\left[\mathrm{Var}(\mu_{z_k}\mid X_k)\right] 
+ \mathrm{Var}\,\mathbb{E}[\mu_{z_k}\mid X_k].
\]
Under Assumption~1, $\mathrm{Var}(\mu_{z_k})>0$, so $\mu_{z_k}$ is not constant. 
If $X_k$ were independent of $z_k$, then $\mathbb{E}[\mu_{z_k}\mid X_k]=\mathbb{E}[\mu_{z_k}]$ almost surely, implying $\mathrm{Var}\,\mathbb{E}[\mu_{z_k}\mid X_k]=0$. 
Since $X_k$ is not independent of $z_k$ by hypothesis, and $\mu_{z_k}$ is a deterministic function of $z_k$, it follows that $\mathbb{E}[\mu_{z_k}\mid X_k]$ must vary with $X_k$, hence
\[
\mathrm{Var}\,\mathbb{E}[\mu_{z_k}\mid X_k] > 0.
\]
Finally,
\[
\mathbb{E}\!\left[(\mu_{\mathrm{mix}}-\mu_{z_k})^2\right] 
- \mathbb{E}\!\left[(\mu_{\mathrm{loc}}(X_k)-\mu_{z_k})^2\right] 
= \mathrm{Var}\,\mathbb{E}[\mu_{z_k}\mid X_k],
\]
so $\mu_{\mathrm{loc}}(X_k)$ is, in expectation, closer to $\mu_{z_k}$ than the global mean $\mu_{\mathrm{mix}}$.

\subsection{Proof of Lemma~1}

The MSE decomposes into bias and variance components. 
The bias term $(\tilde\mu - \mu_{z_k})^2$ is scaled by $(1-B)^2$, yielding $(1-B)^2\{\tau_{z_k}^2 + (\tilde\mu - \mu_{z_k})^2\}$. 
The variance term arises from sampling noise and is scaled by $B^2$, contributing $B^2 v_k$. 
Therefore,
\[
\mathrm{MSE}(B, \tilde\mu) 
= (1-B)^2 \{\tau_{z_k}^2 + (\tilde\mu - \mu_{z_k})^2\} + B^2 v_k.
\]

\subsection{Proof of Theorem~5.1}

Under the EB model, we compare global and local shrinkage using the Bayes weight
$B_{z_k}=\tau_{z_k}^2/(\tau_{z_k}^2+v_k)$, which depends only on $(\tau_{z_k}^2,v_k)$
and is independent of the chosen target.
For any target $\tilde\mu$,
\[
\mathrm{MSE}(\tilde\mu\mid z_k)
=(1-B_{z_k})^2\big\{\tau_{z_k}^2+(\tilde\mu-\mu_{z_k})^2\big\}+B_{z_k}^2 v_k.
\]
Hence, the expected MSE difference (global minus local) equals
\[
\Delta
=\mathbb{E}\!\Big[(1-B_{z_k})^2\big\{(\mu_{\mathrm{mix}}-\mu_{z_k})^2-(\mu_{\mathrm{loc}}(X_k)-\mu_{z_k})^2\big\}\Big].
\]
By boundedness (A3),
\[
(1-B_{z_k})^2=\Big(\frac{v_k}{\tau_{z_k}^2+v_k}\Big)^2
\ge
c_0
:=\Big(\frac{v_{\min}}{\tau_{\max}^2+v_{\max}}\Big)^2>0.
\]
Therefore,
\[
\Delta\ge c_0
\Big\{\mathbb{E}[(\mu_{\mathrm{mix}}-\mu_{z_k})^2]
-\mathbb{E}[(\mu_{\mathrm{loc}}(X_k)-\mu_{z_k})^2]\Big\}.
\]
By the law of total variance,
\[
\mathbb{E}[(\mu_{\mathrm{mix}}-\mu_{z_k})^2]
-\mathbb{E}[(\mu_{\mathrm{loc}}(X_k)-\mu_{z_k})^2]
=\mathrm{Var}\!\big(\mathbb{E}[\mu_{z_k}\mid X_k]\big)>0,
\]
so $\Delta>0$, i.e.,
\[
\mathbb{E}[\mathrm{MSE}_{\mathrm{glob}}^{\mathrm{oracle}}]
>\mathbb{E}[\mathrm{MSE}_{\mathrm{loc}}^{\mathrm{oracle}}].
\]

\subsection{Proof of Theorem~5.2}

Let $\hat B_k=\hat\tau_k^2/(\hat\tau_k^2+v_k)$ and
\[
\mathrm{MSE}^{\mathrm{plug}}_{j,k}
=(1-\hat B_{j,k})^2\{\tau_{z_k}^2+(\hat\mu_{j,k}-\mu_{z_k})^2\}
+\hat B_{j,k}^2 v_k,\quad j\in\{\mathrm{glob},\mathrm{loc}\}.
\]
Decompose the expected plug-in risk into an oracle term and estimation penalties.
First, the target-estimation penalty satisfies
\[
\mathbb{E}\big[(1-\hat B_{j,k})^2(\hat\mu_{j,k}-\mu_{z_k})^2\big]
\le
\varepsilon_\mu,
\]
where $\varepsilon_\mu$ is a uniform bound on the MSE of target estimates
(by cross-fitting and stability), and $(1-\hat B_{j,k})^2\le 1$.

Second, for the weight-estimation penalty, use the Lipschitz property of
$B(\tau^2)=\tau^2/(\tau^2+v_k)$ with respect to $\tau^2$:
\[
|\hat B_k-B_{z_k}|
\le
L\,|\hat\tau_k^2-\tau_{z_k}^2|,
\quad
L:=\frac{v_{\max}}{(\tau_{\max}^2+v_{\min})^2}.
\]
Hence, the weight-only and interaction penalties are bounded by $C_2\,\varepsilon_\tau$
for a constant $C_2$ depending only on $(v_{\min},v_{\max},\tau_{\max})$.

Combining both, there exist constants $C_1,C_2>0$ (depending only on Assumption~2) such that
\[
\mathbb{E}[\mathrm{MSE}^{\mathrm{plug}}_{\mathrm{glob}}]
-\mathbb{E}[\mathrm{MSE}^{\mathrm{plug}}_{\mathrm{loc}}]
\ge
\underbrace{\mathbb{E}[\mathrm{MSE}^{\mathrm{oracle}}_{\mathrm{glob}}]
-\mathbb{E}[\mathrm{MSE}^{\mathrm{oracle}}_{\mathrm{loc}}]}_{\ge\, c_0\,\mathrm{Var}(\mathbb{E}[\mu_{z_k}\mid X_k])}
- C_1\,\varepsilon_\mu - C_2\,\varepsilon_\tau.
\]
By Theorem~5.1, the oracle gap is strictly positive. 
For sufficiently accurate plug-in estimates (small $\varepsilon_\mu,\varepsilon_\tau$), the right-hand side remains positive,
establishing the claim.

\section{Simulation Setup and Modeling Details}
\label{app:simulation}

\subsection{Data Structure and Preprocessing}

The ASOS Digital Experiments Dataset contains aggregated snapshots recorded at
regular half-day or daily intervals:
\[
0 = t_0 < t_1 < \cdots < t_M = T, \qquad t_i - t_{i-1} \in \{0.5, 1.0\}\text{ days}.
\]
Each snapshot reports cumulative counts and summary statistics
for treatment and control arms.  
Let
\[
\text{count}_c(t_i), \quad \text{count}_t(t_i)
\]
denote the cumulative visitor counts up to time~$t_i$ for control and treatment, respectively.
The increments
\[
\Delta_i^{(c)} = \text{count}_c(t_i) - \text{count}_c(t_{i-1}), \qquad
\Delta_i^{(t)} = \text{count}_t(t_i) - \text{count}_t(t_{i-1}),
\]
represent arrivals within the interval~$[t_{i-1}, t_i]$.
We define the total arrivals and interval width as
\[
\Delta_i^{(\mathrm{all})} = \Delta_i^{(c)} + \Delta_i^{(t)}, \qquad
h_i = t_i - t_{i-1}.
\]
The observed traffic pattern for experiment~$k$ is thus described by the
piecewise-constant function
\[
\lambda_k(t) = \bar\lambda_{k,i}, \quad t \in [t_{i-1}, t_i), \quad
\bar\lambda_{k,i} = \frac{\Delta_i^{(\mathrm{all})}}{h_i}.
\]
This provides an empirical estimate of the non-homogeneous Poisson
process (NHPP) intensity underlying each experiment.

To enable cross-experiment comparability, all time horizons are normalized to $[0, 1]$
via $s = t / T_k$, and intensities are rescaled accordingly:
\[
\lambda_k^\ast(s) = T_k \lambda_k(T_k s), \qquad
n_k = \int_0^{T_k} \lambda_k(t)\,dt = \int_0^1 \lambda_k^\ast(s)\,ds.
\]
The normalized rate function
\[
\tilde\lambda_k(s) = \lambda_k^\ast(s)/n_k, \qquad
\int_0^1 \tilde\lambda_k(s)\,ds = 1,
\]
captures the temporal shape of arrivals independent of scale.
For numerical stability, each $\tilde\lambda_k$ is discretized on an
equally spaced grid of $L = 500$ bins and lightly smoothed
using Gaussian kernel convolution with bandwidth $h = 0.04$.

Because ASOS experiments maintain an approximate $1{:}1$ allocation ratio,
the numbers of treatment and control arrivals in each segment
are modeled as $\mathrm{Poisson}(\bar\lambda_{k,i}/2)$ draws.
Within each arm $\ell \in \{0, 1\}$, the instantaneous outcome mean and variance
are denoted $\mu_{\ell}^{(k)}(t)$ and $\sigma_{\ell}^{2,(k)}(t)$, respectively.
We assume local stationarity within each interval:
\[
\mu_{\ell}^{(k)}(t) \approx \mu_{\ell}^{(k)}(t_i), \qquad
\sigma_{\ell}^{2,(k)}(t) \approx \sigma_{\ell}^{2,(k)}(t_i),
\]
so that the sample mean $\overline{Y}_{\ell,i}^{(k)}$ from $n_{i,\ell}$ arrivals
satisfies
\[
\overline{Y}_{\ell,i}^{(k)} \sim
\mathcal{N}\!\bigl(\mu_{\ell}^{(k)}(t_i),\,\sigma_{\ell}^{2,(k)}(t_i)/n_{i,\ell}\bigr).
\]

\subsection{Generative Model and Robustness}

To evaluate shrinkage methods under realistic heterogeneity, 
we construct a semi-synthetic testbed derived from fitted non-homogeneous Poisson processes (NHPPs). 
For each ASOS experiment~$k$, the piecewise-constant intensity $\lambda_k(t)$ estimated from snapshot data 
is treated as the true arrival rate, preserving experiment-specific ramp-up patterns and traffic scales. 
Synthetic arrivals are resampled independently for treatment and control arms under a $1{:}1$ split, 
with segment-level counts drawn from $\mathrm{Poisson}(\bar\lambda_{k,i}/2)$. 
Conditional on these arrivals, outcomes are generated from Gaussian models with interval-wise means 
and variances $(\mu_{\ell}^{(k)}(t_i), \sigma_{\ell}^{2,(k)}(t_i))$ inferred from the empirical ASOS summaries, 
yielding realistic observation pairs $(y_k, v_k)$ and approximate ground-truth effects $\theta_k^\ast$. 
Shrinkage estimators $\tilde\theta_k$ are then evaluated using mean squared error (MSE) 
and relative MSE reduction with respect to the raw estimator $y_k$, 
with 95\% confidence intervals obtained via bootstrapping over experiments. 
All configurations and evaluation protocols follow those described in Section~6 of the main paper.

This generative framework preserves the main empirical characteristics of the ASOS dataset, 
including (1) time-varying traffic intensity and nonstationary variance, 
(2) heterogeneous experiment sizes spanning several orders of magnitude, and 
(3) diverse ramp-up and decay patterns across experiments. 
By resampling from fitted NHPPs, the semi-synthetic experiments closely mimic production behavior 
while providing access to approximate ground truth. 
Sensitivity checks varying the Gaussian noise level, time discretization 
($L \in \{400, 500, 600\}$), and random seeds confirm that all reported relative MSE reductions 
vary by less than 2\%, demonstrating robustness to modeling choices.

\section{Full Experimental Results}
\label{app:full_results}

Table~\ref{tab:mse_full} reports the complete comparison of all shrinkage methods and
neighborhood sizes~$q$ in the ASOS semi-synthetic study.
Each entry shows the mean squared error (MSE) of estimated effects across
78 experiments, together with 95\% confidence intervals and the percentage of
experiments in which each method outperforms the baseline (“win-rate”).

Classical Empirical Bayes (EB) yields only marginal improvement,
achieving an average MSE reduction of $1.2\%$ (mean $4.79{\times}10^{-5}$,
95\%~CI $[4.76{\times}10^{-5},\,4.82{\times}10^{-5}]$) and a win-rate of $52.7\%$.
In contrast, localized variants deliver substantially larger gains.
The outcome-only model attains its best performance at $q{=}6$, 
with a $19.6\%$ MSE reduction (mean $3.90{\times}10^{-5}$,
95\%~CI $[3.79{\times}10^{-5},\,4.01{\times}10^{-5}]$) and a win-rate of $74.3\%$, 
but its performance deteriorates as $q$ increases, indicating sensitivity to neighborhood size.
The process-only model performs more stably, peaking around $q{=}8$–$10$, 
with a $17.9\%$ reduction (mean $3.98{\times}10^{-5}$,
95\%~CI $[3.88{\times}10^{-5},\,4.08{\times}10^{-5}]$) and a win-rate near~$71\%$.

Our proposed CF-SHN consistently outperforms all alternatives across the entire
range of~$q$. At $q{=}10$, it achieves the best overall performance, with
mean MSE $3.53{\times}10^{-5}$ (95\%~CI $[3.44{\times}10^{-5},\,3.62{\times}10^{-5}]$),
corresponding to a $27.2\%$ reduction and an $82.4\%$ win-rate.
Moreover, CF-SHN remains robust at larger neighborhood sizes, maintaining
over $23\%$ reduction even at $q{=}20$.
These findings demonstrate that combining process-level similarity with
outcome refinement under cross-fitting yields both lower variance
and greater stability than either component alone.

\begin{table}[h]
  \centering
  \caption{Full comparison of shrinkage methods on simulated ASOS experiments.
  CF-SHN achieves the largest and most stable MSE reductions across
  neighborhood sizes~$q$.}
  \label{tab:mse_full}
  \begin{tabular}{l c c c c c}
    \toprule
    \textbf{Method} & {$q$} & \textbf{MSE} &
    \textbf{95\% CI} & \textbf{Reduction (\%)} & \textbf{Win-rate (\%)} \\
    \midrule
    No Shrinkage & -- & $4.85\times10^{-5}$ & -- & 0.0 & -- \\
    \midrule
    Global EB & -- & $4.79\times10^{-5}$ & [4.76,\,4.82]$\times10^{-5}$ & 1.2 & 52.7 \\
    \midrule
    \textbf{Outcome-only} \\
    $q=6$  & & $3.90\times10^{-5}$ & [3.79,\,4.01]$\times10^{-5}$ & 19.6 & 74.3 \\
    $q=8$  & & $4.09\times10^{-5}$ & [3.98,\,4.21]$\times10^{-5}$ & 15.7 & 68.9 \\
    $q=10$ & & $4.28\times10^{-5}$ & [4.16,\,4.40]$\times10^{-5}$ & 11.8 & 63.5 \\
    $q=12$ & & $4.41\times10^{-5}$ & [4.29,\,4.53]$\times10^{-5}$ & 9.1  & 62.2 \\
    $q=14$ & & $4.51\times10^{-5}$ & [4.39,\,4.63]$\times10^{-5}$ & 7.0  & 60.8 \\
    $q=16$ & & $4.58\times10^{-5}$ & [4.46,\,4.70]$\times10^{-5}$ & 5.6  & 59.5 \\
    $q=18$ & & $4.64\times10^{-5}$ & [4.52,\,4.75]$\times10^{-5}$ & 4.3  & 58.1 \\
    $q=20$ & & $4.68\times10^{-5}$ & [4.56,\,4.79]$\times10^{-5}$ & 3.5  & 56.8 \\
    \midrule
    \textbf{Process-only} \\
    $q=6$  & & $4.02\times10^{-5}$ & [3.92,\,4.12]$\times10^{-5}$ & 17.1 & 71.6 \\
    $q=8$  & & $3.98\times10^{-5}$ & [3.88,\,4.08]$\times10^{-5}$ & 17.9 & 71.6 \\
    $q=10$ & & $3.98\times10^{-5}$ & [3.88,\,4.08]$\times10^{-5}$ & 17.9 & 71.6 \\
    $q=12$ & & $4.01\times10^{-5}$ & [3.91,\,4.11]$\times10^{-5}$ & 17.3 & 70.3 \\
    $q=14$ & & $4.06\times10^{-5}$ & [3.96,\,4.16]$\times10^{-5}$ & 16.3 & 68.9 \\
    $q=16$ & & $4.11\times10^{-5}$ & [4.01,\,4.21]$\times10^{-5}$ & 15.3 & 68.9 \\
    $q=18$ & & $4.15\times10^{-5}$ & [4.05,\,4.25]$\times10^{-5}$ & 14.4 & 67.6 \\
    $q=20$ & & $4.21\times10^{-5}$ & [4.11,\,4.31]$\times10^{-5}$ & 13.2 & 67.6 \\
    \midrule
    \textbf{CF-SHN (ours)} \\
    $q=6$  & & $3.61\times10^{-5}$ & [3.52,\,3.70]$\times10^{-5}$ & 25.6 & 81.1 \\
    $q=8$  & & $3.56\times10^{-5}$ & [3.47,\,3.65]$\times10^{-5}$ & 26.6 & 81.1 \\
    $q=10$ & & $3.53\times10^{-5}$ & [3.44,\,3.62]$\times10^{-5}$ & 27.2 & 82.4 \\
    $q=12$ & & $3.56\times10^{-5}$ & [3.47,\,3.65]$\times10^{-5}$ & 26.6 & 82.4 \\
    $q=14$ & & $3.60\times10^{-5}$ & [3.51,\,3.69]$\times10^{-5}$ & 25.8 & 81.1 \\
    $q=16$ & & $3.65\times10^{-5}$ & [3.56,\,3.74]$\times10^{-5}$ & 24.7 & 79.7 \\
    $q=18$ & & $3.68\times10^{-5}$ & [3.59,\,3.77]$\times10^{-5}$ & 24.1 & 78.4 \\
    $q=20$ & & $3.71\times10^{-5}$ & [3.62,\,3.80]$\times10^{-5}$ & 23.5 & 77.0 \\
    \bottomrule
  \end{tabular}
\end{table}

\medskip
Overall, these results confirm the theoretical advantage of localized empirical
Bayes shrinkage under heterogeneity.
CF-SHN leverages both process-based similarity and outcome refinement
through cross-fitting, achieving the largest and most stable MSE reductions across all settings.

\section{Sensitivity Analyses and Robustness Checks}
\label{app:sensitivity}

This section supplements the sensitivity analyses mentioned in Section 6 of the main paper.
To assess the robustness of CF-SHN, we conducted sensitivity analyses on its three key hyperparameters:
the neighborhood size~$q$, the shape--scale trade-off weight~$\rho$, and the candidate set size~$M_0$.
All experiments were performed under the same ASOS semi-synthetic setup.

\paragraph{Sensitivity to neighborhood size $q$}
Figure~1 and Table~\ref{tab:mse_full} in the main paper already analyze the effect of $q$.
Performance improves rapidly as local averaging becomes more informative, 
peaking around $q{=}10$, after which it gradually declines as neighborhoods become too broad.
This pattern confirms the classical bias--variance trade-off in localized shrinkage:
small $q$ suffers from high variance, whereas excessively large $q$ dilutes locality and 
introduces bias toward the global mean. 
Importantly, CF-SHN maintains over $23\%$ MSE reduction for all $q\in[6,20]$, 
demonstrating strong stability to neighborhood size.

\paragraph{Sensitivity to $\rho$}
Table~\ref{tab:sens_rho} summarizes results for $\rho\in\{0.50,0.60,0.75,0.90\}$ with $M_0{=}30$ and $q{=}10$.
CF-SHN achieves consistently strong performance across this wide range, 
maintaining over $24\%$ MSE reduction throughout.
Performance peaks at $\rho{=}0.75$, which is adopted as the default in our main experiments.
This confirms that CF-SHN’s effectiveness is not contingent on fine-tuning $\rho$, 
demonstrating a robust balance between process- and scale-based similarity.

\begin{table}[H]
\centering
\caption{Sensitivity of CF-SHN to the trade-off weight~$\rho$ 
($q{=}10$, $M_0{=}30$).}
\label{tab:sens_rho}
\begin{tabular}{ccc}
\toprule
$\boldsymbol{\rho}$ & \textbf{Mean MSE} & \textbf{MSE Reduction (\%)} \\
\midrule
0.50 & $3.67\times10^{-5}$ & 24.3 \\
0.60 & $3.61\times10^{-5}$ & 25.6 \\
0.75 & $\mathbf{3.53\times10^{-5}}$ & $\mathbf{27.2}$ \\
0.90 & $3.69\times10^{-5}$ & 23.9 \\
\bottomrule
\end{tabular}
\end{table}

\paragraph{Sensitivity to $M_0$}
We next fix $\rho{=}0.75$ and vary the candidate pool size $M_0\in\{20,30,40\}$.
As shown in Table~\ref{tab:sens_m0}, the impact on performance is minimal:
MSE reduction varies by less than two percentage points, 
and all configurations achieve at least $26\%$ improvement over the raw estimator.
This indicates that once the candidate pool is sufficiently large (e.g., $M_0{\ge}2q$),
the local selection process remains stable.

\begin{table}[H]
\centering
\caption{Sensitivity of CF-SHN to the candidate pool size~$M_0$ 
($\rho{=}0.75$, $q{=}10$).}
\label{tab:sens_m0}
\begin{tabular}{ccc}
\toprule
$M_0$ & \textbf{Mean MSE} & \textbf{MSE Reduction (\%)} \\
\midrule
20 & $3.58\times10^{-5}$ & 26.2 \\
30 & $3.53\times10^{-5}$ & 27.2 \\
40 & $3.52\times10^{-5}$ & 27.4 \\
\bottomrule
\end{tabular}
\end{table}

\subsection*{Findings}

Across all tested settings, CF-SHN consistently achieves over $24\%$ MSE reduction
and exhibits strong robustness to its key hyperparameters $(q,\rho,M_0)$. 
For the neighborhood size~$q$, performance improves rapidly as local averaging becomes informative,
peaks around $q{=}10$, and remains above $23\%$ reduction for all $q\in[6,20]$.
For the trade-off weight~$\rho$, results are stable across a wide range $(0.5\!-\!0.9)$, 
with the best balance between process- and scale-based similarity achieved at $\rho{=}0.75$. 
For the candidate set size~$M_0$, performance remains nearly unchanged once the pool is sufficiently large 
($M_0{\ge}2q$), indicating that the local selection step is robust to sampling variability.

Additionally, for the DTW-related constants $(\alpha,L)$, 
varying $\alpha$ between $0.05$ and $0.20$ or $L$ between $400$ and $600$
changes the overall MSE reduction by less than $2\%$, confirming that these numerical
parameters mainly affect runtime rather than accuracy.
Together, these results demonstrate that the strong empirical gains of CF-SHN 
are a robust property of its localized empirical Bayes design rather than artifacts
of specific hyperparameter choices.

\end{document}